\documentclass[]{interact}

\usepackage{epstopdf}
\usepackage[caption=false]{subfig}

\usepackage[numbers,sort&compress]{natbib}
\bibpunct[, ]{[}{]}{,}{n}{,}{,}

\theoremstyle{plain}

\theoremstyle{definition}

\theoremstyle{remark}

\usepackage{graphicx}

\usepackage{amssymb}

\usepackage{amsbsy}
\usepackage{amsmath}
\usepackage{multirow}
\usepackage[table,xcdraw]{xcolor}
\usepackage{graphicx,rotating,booktabs}
\usepackage[export]{adjustbox}

\begin{document}

\articletype{Original Research Article}

\title{Empirical Likelihood Estimation for Linear Regression Models with AR(p) Error Terms}

\author{
\name{\c{S}enay \"{O}zdemir \textsuperscript{a}\thanks{CONTACT \c{S}. \"{O}zdemir. Email: senayozdemir@aku.edu.tr},
 Ye\c{s}im G\"{u}ney \textsuperscript{b},       
 Yetkin Tua\c{c}\textsuperscript{b}             
  and       Olcay Arslan\textsuperscript{b}}
\affil{\textsuperscript{a}Department of Statistics, Afyon Kocatepe University, 03200, Afyonkarahisar, Turkey ;\\
\textsuperscript{b}Department of Statistics, Ankara University, 06100, Ankara, Turkey}
}

\maketitle

\begin{abstract}
Linear regression models are useful statistical tools to analyze data sets in several  different  fields. There are several methods to estimate the parameters of a linear regression model. These methods usually perform under normally distributed and uncorrelated errors with zero mean and constant variance. However, for some data sets error terms may not satisfy these or some of these assumptions.  If  error terms are correlated, such as  the  regression models with autoregressive (AR(p)) error terms, the Conditional Maximum Likelihood (CML) under normality assumption or the Least Square (LS)  methods are often used to estimate the  parameters  of  interest. For  CML estimation a distributional assumption on  error terms is  needed to  carry on estimation, but, in practice,  such distributional assumptions  on  error terms  may not be plausible.  Therefore,  in  such  cases some alternative  distribution free methods  are  needed to conduct the parameter  estimation. In this paper, we propose to estimate the parameters of a linear regression model with AR(p) error term using the Empirical Likelihood (EL) method, which  is  one  of  the  distribution  free estimation  methods.  A small simulation study and a numerical example are provided to evaluate the performance of the proposed estimation method over the CML method. The results of simulation study  show  that  the   proposed estimators based  on  EL method are  remarkably  better than the  estimators obtained from  the CML method  in terms of  mean squared  errors (MSE)   and  bias  in  almost  all  the  simulation  configurations.   These findings  are  also   confirmed  by  the  results  of  the  numerical and real data examples.
\end{abstract}


\section{Introduction}
\label{intro}
Consider the following linear regression model

\begin{equation}\label{1}
    y_{t}=\boldsymbol{{x_{t}^{T}}{\beta}}+\varepsilon_{t} \quad \textrm{for}\quad t=1,2,\ldots,N
\end{equation}

\noindent where $y_{t}$ are the $t-th$ response variable, $\boldsymbol{x_{t}}\in R^{M}$ are the design vector, $\boldsymbol\beta \in R^M$ is the  unknown M-dimensional parameter vector and $\varepsilon_{t}$ are the uncorrelated  errors with $E(\varepsilon_{t})=0$ and $Var(\varepsilon_{t})=\sigma^2$.

It is known that the LS estimators  of the regression parameters are obtained by minimizing the
sum of the squared residuals or solving the following estimating equation

\begin{equation}\label{2}
\frac{1}{N}\sum_{t=1}^{N}\left(y_{t}-\boldsymbol{{x_{t}^{T}}{\beta}}\right)\boldsymbol{x_{t}} =\boldsymbol{0}.
\end{equation}

\noindent The LS estimators are the minimum variance unbiased  estimators  of  $\boldsymbol{\beta}$   if $\varepsilon_{t}$ are normally distributed. However, in  real  data applications  the normality assumption may not be   completely satisfied. If   the  normally  distributed  error  term  is  not a  reasonable  assumption  for a  regression  model, some    alternative distributions  can  be  used   as   the error  distribution and  the   maximum likelihood  estimation  method   can be applied   to  estimate  the  regression  and   other  model  parameters. On  the  other  hand, if   an  error  distribution is  not be easy  to  specify   some  alternative distribution  free  estimation  methods  should  be  preferred   to  obtain    estimators  for  the  parameters  of  interest.  One  of  such  methods  is   the EL  method introduced by Owen \cite{Owen1988,Owen1990}.   A noticeable advantage of EL method  is  that  it creates likelihood-type inference without specifying any  distributional  model for the data. The EL method supposes that  there are  unknown  $\pi_{t}$,  $t=1,2,\ldots,N$,  probability weights for each observation  and it tries to estimate  these probability weights by maximizing an EL function defined as the production of $\pi_{t}$s under some constraints related with $\pi_{t}$s, and the regression  parameters.  The  EL method can be mathematically defined as follows. Maximize the following EL function

\begin{equation}\label{3}
L(\boldsymbol\beta)=\prod_{t=1}^{N}\pi_{t}
\end{equation}

\noindent under the constraints
\begin{equation}\label{4}
\sum_{t=1}^{N}\pi_{t}=1
\end{equation}
\begin{equation}\label{5}
\sum_{t=1}^{N}\pi_{t}\boldsymbol{x_{t}}\left(y_{t}-\boldsymbol{x_{t}^{T}\beta}\right)=\boldsymbol{0}.
\end{equation}

\noindent Note that the constraint given in equation  (\ref{5}) is similar to the estimating equation given in equation (\ref{2}). The only difference is that in equation (\ref{2}) we use the equal known weight $(1/n)$ for each observation, but in equation (\ref{5}) we use the unknown probability weights $\pi_{t}$s by considering that each observation has different contribution to the estimation procedure. Further, if  we  also  want  to  estimate  the  error  variance  along  with the  regression  parameters   we  add  the  following constraint to  the  constraints  given  in  equations  (\ref{4})-(\ref{5}).

\begin{equation}\label{6}
\sum_{t=1}^{N}\pi_{t}\left[(y_{t}-\boldsymbol{x_{t}^{T}\beta})^{2}-\sigma^{2}\right]=0.
\end{equation}

Since the EL method can be used for estimating parameters, constructing confidence regions  and  testing statistical hypothesis, it is a useful tool for making statistical inference when it is not too easy to assign a distribution to data. There are several remarkable studies on the EL method after it was first introduced by Owen \cite{Owen1988,Owen1990,Owen1991}. Some of these papers can be summarized as follows. Hall and La Scala\cite{Hall1990} studied on main features of the EL, Kolaczyk \cite{Kolaczyk1994} adapted it in generalized linear regression model,Qin and Lawless \cite{Qin1994} combined general estimating equations and the EL, Chen et al. \cite{Chen1993,Chen1994,Chen1996,Chen2003,Chen2009} considered this method for constructing confidence regions  and parameter estimation with additional constraints,Newey and Smith\cite{Newey2004} studied higher-order properties  generalized methods of moments and generalized empirical likelihood estimators, Shi and Lau \cite{Shi1999}   considered  for robustifying the constraint in equation (\ref{5}) using  median constraint, and Bondell and Stefanski \cite{Bondell2013} suggested a robust estimator by maximizing a generalized EL function instead of the EL function given in equation (\ref{3}). Recently, \"{O}zdemir and Arslan \cite{Ozdemir2018b} have considered  using constraints based on robust M estimation in EL estimation method. Also, \"{O}zdemir and Arslan \cite{Ozdemir2018a} have proposed an alternative algorithm to compute  EL estimators.

The estimation of $\pi_{t}$s and hence the model parameters $\boldsymbol{\beta}$ and $\sigma^{2}$ can be done by maximizing the EL function given in equation (\ref{3}) under the constraints (\ref{4})-(\ref{6}).  In general, Lagrange multipliers method can be used for these types of constrained optimization problems. For our problem the Lagrange function will be  as follows

\begin{eqnarray*}
   L(\boldsymbol{\pi,\beta},\lambda_{0},\boldsymbol\lambda_{1}^{T},\lambda_{2})&=&\sum_{t=1}^{N}log(n\pi_{t})+\lambda_{0}\left(\sum_{t=1}^{N}\pi_{t} - 1\right)\\
   &+&\boldsymbol\lambda_{1}^{T}\sum_{t=1}^{N} \pi_{t} \boldsymbol{x_{t}}\left(y_{t}-\boldsymbol{x_{t}^{T}\beta}\right)\\
   &+&\lambda_2\sum_{t=1}^{N} \pi_{t}\left[(y_{t}-\boldsymbol{x_{t}^{T}\beta})^{2}-\sigma^{2}\right]
\end{eqnarray*}

\noindent where $\boldsymbol{\pi}=[\pi_{1},\pi_{2},\ldots,\pi_{N}]^{T}$, $\lambda_{0},\lambda_{2} \in R^1$ and $\boldsymbol\lambda_{1}^{T}\in R^{M}$ are Lagrange multipliers. Taking the derivatives of Lagrange function with respect to each $\pi_{t}$ and setting to zero we get
\begin{equation}\label{7}
    \pi_{t}=\frac{1}{N+\boldsymbol\lambda_{1}^{T}\boldsymbol{x_{t}}\left(y_{t}-\boldsymbol{x_{t}^{T}\beta}\right)+\lambda_2\left[(y_{t}-\boldsymbol{x_{t}^{T}\beta})^{2}-\sigma^{2}\right] }.
\end{equation}

\noindent Substituting $\pi_{t}$ given in equation (\ref{7}) in the EL function and constrains, the optimization problem is reduced to the problem of finding $\boldsymbol\beta$,  $\sigma$  and Lagrange multipliers. However, this problem is still not easy to handle to obtain the estimators  for $\boldsymbol\beta$ and $\sigma$.   The solution of this problem are considered by several author using several different approaches. For details of the algorithms  they  are suggesting, one can see the papers \cite{Owen1988,Owen1990,Owen1991,Owen2001,Owen2013,Yang2013,Ozdemir2018b}.

It should be noted that in all of the mentioned papers researchers consider   the EL estimation method to  estimate the parameters of  a regression  model   with  uncorrelated  error terms.  However, in practice, uncorrelated error assumption may not be plausible for some data sets. For these data sets  regression analysis should be carried on with an autoregressive error terms regression model (regression  model  with AR(p) error  terms).  An  autoregressive  error terms  regression  model   is  defined as follows.

\begin{equation}\label{8}
    y_{t}=\sum_{i=1}^{M}{x_{t,i}\beta_{i}+e_{t}} , t=1,2,\dots,N,
\end{equation}

\noindent where $y_{t}$ is the response variable, $x_{t,i}$ is the predictor variable, $\beta_{i}$ is the unknown regression parameter and $e_{t}$ is the  AR(p)  error term with

\begin{equation}\label{9}
    e_{t}=\phi_{1} e_{t-1}+\dots+\phi_{p} e_{t-p}+a_{t}.
\end{equation}

\noindent Here $\phi_{j}, j=1,2,3,...,p$ are the unknown autoregressive parameters and $a_{t}, t=1,2,3,...,N$ are  i.i.d. random variables with $E(a_{t})=0$ and $Var(a_{t})=\sigma^{2}$. Note that this regression equation is different from the regression equation given in (\ref{1}). However, using the back shift operator $B$, this equation can be transformed  to the usual regression  equation as follows. Let
\begin{equation*}
a_{t}=\Phi(B) e_{t}= e_{t}-\phi_{1} e_{t-1}-\dots-\phi_{p} e_{t-p},
\end{equation*}

\begin{equation}\label{10}
   \Phi(B)y_{t}=y_{t}-\phi_{1}y_{t-1}-\dots-\phi_{p}y_{t-p},
\end{equation}

\noindent and

\begin{equation}\label{11}
   \boldsymbol{\Phi(B)x_{t}}=\boldsymbol{x_{t}}-\phi_{1}\boldsymbol{x_{t-1,i}}-\dots-\phi_{p}\boldsymbol{x_{t-p,i}}.
\end{equation}
\\
\noindent Then, the regression model given in (\ref{8}) can be rewritten as

\begin{equation}\label{12}
   \Phi(B)y_{t}=\sum_{i=1}^{M}{\beta_{i}\Phi(B)x_{t,i}+a_{t}} , t=1,2,\dots,N.
\end{equation}

In  literature, parameters of  an  autoregressive error term  regression  model are  estimated  using   LS,  ML or CML estimation methods.   Some of the related papers are    \cite{Alpuim2008},   \cite{Beach1978},  \cite{Tiku1999},  \cite{Tuac2018} and \cite{Tuac2019}. In all of the mentioned papers some known distributions,  such as normal or t, are assumed as the error distribution to carry on estimation of the parameters of interest in this model.  However, since imposing appropriate  distributional  assumptions  on the error  term of  a  regression  model  may  not be easy   some  other  distribution free  estimation  methods may  be  preferred  to  carry on  regression  analysis  of  a  data set.  In this study, unlike the papers  in  literature, we will not assume any distribution for the error terms and propose to use the EL estimation method to estimate the parameters of the linear regression model described in the previous paragraph.

The rest of the paper is organized as follows. In Section 2, the CML and the EL estimation methods for the linear models with AR(p) error terms are given. In  Section 3, A small simulation study   and  a   numerical  and a real data examples  are  provided  to  assess the performance  of  the EL   based   estimators  over  the  estimators obtained  from  the  classical  CML  method.   Finally,  we draw some conclusions in Section 4.

\section{Parameter Estimation for Linear Regression Models with AR(p) Error Terms }
\label{Sec2}
In this section, we describe in detail how  the  EL  method  is  used  to  estimate  the  parameters  of  an  autoregressive error term  regression  model.    We   will first  recall   the   CML method   under normality  assumption  of   $a_t$.  Note  that since the exact likelihood function could be well approximated by the conditional likelihood function \cite{Ansley1979} CML estimation method are often used  in
cases where ML estimation method is  not feasible  to carry on.

\subsection{Conditional Maximum Likelihood Estimation}
\label{SubSec2.1}
In general, a system of nonlinear equations of the parameters have to be solved to obtain ML  estimators. However,  since in  most  of  the   cases  ML  estimators cannot  be    analytically  obtained    numerical procedures should   be  used to  get  the estimates  for  the  parameters of  interest. An alternative way for numerical maximization of the exact likelihood function is to regard the value of the first $p$ observations as known and to maximize the likelihood function conditioned on the first observations. In this part, the CML estimation method will be considered for the regression model given in (\ref{8}).

Let the error terms $a_{t}$  in  the  regression  model  given  in  equation (\ref{12}) have normal distribution with zero mean and $\sigma^2$ variance.
Then,  the conditional log-likelihood function will be  as

\begin{equation}\label{13}
\ln L=c-\frac{N-p}{2}\ln \sigma^{2}-\frac{1}{2\sigma ^{2}}
\sum_{t=p+1}^{N}\left( \Phi \left( B\right) y_{t}-\sum_{i=1}^{M}\beta _{i}\Phi \left( B\right) x_{t,i}\right)^{2}
\end{equation}


\noindent  \cite{Alpuim2008}. Taking the derivatives of the conditional log-likelihood function with
respect to the unknown parameters, setting them to zero  and rearranging the resulting   equations  yields   the following
estimating equations  for  the  unknown  parameters  of  the  regression  model under  consideration

\begin{equation}\label{14}
\widetilde{\boldsymbol{\beta}}=\left[ \widetilde{\boldsymbol{\Phi }}\left(
B\right) \boldsymbol{X}^{T}\widetilde{\boldsymbol{\Phi }}\left( B\right)
\boldsymbol{X}\right] ^{-1}\left[ \widetilde{\boldsymbol{\Phi }}\left(
B\right) \boldsymbol{X}^{T}\widetilde{\Phi }\left( B\right) Y\right]
\end{equation}

\begin{equation} \label{15}
\widetilde{\sigma}^{2}=
\small{\frac{1}{N-p}\left[\widetilde{\Phi}\left(B\right)
Y-\widetilde{\boldsymbol{\Phi }}\left( B\right) \boldsymbol{X}^{T}\widetilde{\boldsymbol{\beta}}\right] ^{T}\left[ \widetilde{\Phi }\left( B\right) Y-
\widetilde{\boldsymbol{\Phi}}\left(B\right) \boldsymbol{X}^{T}\widetilde{\boldsymbol{\beta}}\right]}
\end{equation}

\begin{equation}\label{16}
\widetilde{\boldsymbol{\Phi}}=\boldsymbol{R}^{-1}(\widetilde{\boldsymbol{\beta}})\boldsymbol{R}_{0}(\widetilde{\boldsymbol{\beta}})
\end{equation}

\noindent where $\widetilde{\boldsymbol{\Phi }}\left( B\right) \boldsymbol{X=}\left[
\widetilde{\Phi }\left( B\right) x_{t,i}\right] $ , $\widetilde{\Phi }\left(
B\right) Y=\left[ \widetilde{\Phi }\left( B\right) y_{t}\right] $,

\begin{eqnarray*}
\boldsymbol{R}\left( \widetilde{\boldsymbol{\beta}}\right)= 
\small{\left[
\begin{array}{cccc}
\sum_{t=p+1}^{N}e_{t-1}^{2} & \sum_{t=p+1}^{N}e_{t-1}e_{t-2}
& \cdots & \sum_{t=p+1}^{N}e_{t-1}e_{t-p} \\
& \sum_{t=p+1}^{N}e_{t-2}^{2} & \cdots & \sum_{t=p+1}^{N}e_{t-2}e_{t-p} \\
\vdots & \vdots & \ddots & \vdots \\
&  & \cdots & \sum_{t=p+1}^{N}e_{t-p}^{2}
\end{array}
\right]}
\end{eqnarray*}

\noindent and
$\boldsymbol{R}_{0}\left( \widetilde{\boldsymbol{\beta}}\right) =\left[
\begin{array}{c}
\sum_{t=p+1}^{N}e_{t}e_{t-1} \\
\sum_{t=p+1}^{N}e_{t}e_{t-2} \\
\vdots \\
\sum_{t=p+1}^{N}e_{t}e_{t-p}%
\end{array}%
\right]. $ \\

\vspace{.1in}

\noindent However, since these  estimating  equations cannot be  explicitly solved  to   get   estimators for the unknown   parameters  some numerical methods should  be  used  to compute  the estimates.   Among   all  numerical methods the   estimating equations suggest   a  simple iteratively  reweighting  algorithm (IRA) to compute estimates  for  the unknown  parameters  \cite{Alpuim2008,Tuac2018}.

\subsection{Empirical Likelihood Estimation}
\label{subSec2.2}

In this section we consider the EL method to estimate the unknown parameters of  the regression  model  given  in  equation (\ref{12}).  The required constraints  related  to  the  parameters   will be formed   similar  to  the EL  estimation  used  in  classical  regression case (uncorrelated error  term  regression  model).  Since  we  will  use  CML  estimation   approach  we will  again  assume  that  first  p  observations  are known  and will  form  the conditional  empirical  likelihood    (CEL)  function  using  the   unknown  probability  weights  $\pi_t$ for the observations   $t=p+1,...,N$.    It  should be  noted  that  in  CML  estimation   case  $a_{t}$  are  assumed  to  have  normal  distribution,  however, in  CEL   estimation  case  we  do  not need  to  assume  any  specific distribution  for $a_{t}$.  Now we can    formulate  the CEL   estimation procedure   as  follows.

Let $\pi_{t}$ for  $t=p+1,...,N$  be the unknown probabilities  for the  observations $i=p+1,...,N$.     Then,   maximize    the  following  CEL  function

\begin{equation}\label{17}
    max_{\pi_{t}\in(0,1)}{\sum_{t=p+1}^{N}log(\pi_{t}) }
\end{equation}

\noindent under the constraints
\begin{equation}\label{18}
    \sum_{t=p+1}^{N}\pi_{t}=1
\end{equation}

\begin{equation}\label{19}
    \sum_{t=p+1}^{N}\pi_{t}\left(\Phi(B)y_{t}-\sum_{i=1}^{M}{\beta_{i}\Phi(B)x_{t,i}}\right)\boldsymbol{\Phi(B)x_{t}}=\boldsymbol{0}
\end{equation}

\begin{equation}\label{20}
    \sum_{t=p+1}^{N}\pi_{t}\left(\Phi(B)y_{t}-\sum_{i=1}^{M}{\beta_{i}\Phi(B)x_{t,i}}\right)\boldsymbol{e_{t,p}}=\boldsymbol{0}
\end{equation}

\begin{equation}\label{21}
    \sum_{t=p+1}^{N}\pi_{t}\left[\left(\Phi(B)y_{t}-\sum_{i=1}^{M}{\beta_{i}\Phi(B)x_{t,i}}\right)^2-\sigma^{2}\right]=0
\end{equation}

\vspace{.1in}

\noindent to  obtain  CEL   estimators for  the  parameters   of  the  regression  model  given in  equations (\ref{12}). Here,\\
$\boldsymbol{\Phi \left( B\right) x_{t}}= \left[ \Phi \left( B\right) x_{t,1},\Phi \left(B\right) x_{t,2},...,\Phi \left( B\right) x_{t,M}\right] ^{T}$  and
$\boldsymbol{e_{t,p}} = \left[ e_{t-1},e_{t-2},...,e_{t-p}\right]^{T}$.

\vspace{.1in}

\noindent Since  this  is a  constraint optimization problem   Lagrange multiplier method  can be used  to solve  it. To  this  extend, let $\lambda_0,\boldsymbol{\lambda}=(\boldsymbol{\lambda}_1^T,\boldsymbol{\lambda}_2^T,\lambda_3)$, $\boldsymbol{\phi}$ and $\boldsymbol{\pi}$ denote the  Lagrange multipliers,  vector  of $\phi_i, i=1,2,...,p$ and  vector  of  $\pi_t, t=p+1,...,N$,   respectively. Then,  the Lagrange function for  this  optimization problem can be written as

\begin{eqnarray*}
  L(\boldsymbol{\beta, \phi},\sigma^{2},\boldsymbol{\pi},\boldsymbol{\lambda},\lambda_0)&=&-\sum_{t=p+1}^{N}{log(\pi_{t})}
  +\lambda_{0}\left(\sum_{t=p+1}^{N}\pi_{t}-1\right)\\
  &+&\boldsymbol\lambda_{1}^{T}\left(\sum_{t=p+1}^{N}\pi_{t}\left(\Phi(B)y_{t}-\sum_{i=1}^{M}{\beta_{i}\Phi(B)x_{t,i}}\right)\boldsymbol{\Phi(B)x_{t}}\right)\\
  &+&\boldsymbol\lambda_{2}^{T}\left(\sum_{t=p+1}^{N}\pi_{t}\left(\Phi(B)y_{t}-\sum_{i=1}^{M}{\beta_{i}\Phi(B)x_{t,i}}\right)\boldsymbol{e_{t,p}}\right)\\
  &+&\lambda_{3}\left(\sum_{t=p+1}^{N}\pi_{t}\left[\left(\Phi(B)y_{t}-\sum_{i=1}^{M}{\beta_{i}\Phi(B)x_{t,i}}\right)^2-\sigma^{2}\right]\right)
\end{eqnarray*}

\vspace{.1in}

\noindent Taking the  derivatives of $L(\boldsymbol{\beta, \phi},\sigma^{2},\boldsymbol{\pi},\boldsymbol{\lambda},\lambda_0)$  with  respect to  $\pi_t$ and the Lagrangian multipliers, and  setting   the  resulting  equations  to  zero  we  get  first  order  conditions  of  this optimization  problem. Solving the  first order conditions with respect  to $\pi_t$  yields

\begin{equation}\label{22}
    \small{\pi_{t}=\frac{1}{(N-p)+\boldsymbol\lambda_{1}^{T}\boldsymbol\Psi_{1,t}+\boldsymbol\lambda_{2}^{T}\boldsymbol\Psi_{2,t}+\lambda_{3}\Psi_{3,t}},
    t=p+1,\dots,N}
\end{equation}

\noindent where

\begin{eqnarray*}
  \boldsymbol\Psi_{1,t} &=& \left(\Phi(B)y_{t}-\sum_{i=1}^{M}{\beta_{i}\Phi(B)x_{t,i}}\right)\boldsymbol{\Phi(B)x_{t}} \\
 \boldsymbol \Psi_{2,t} &=& \left(\Phi(B)y_{t}-\sum_{i=1}^{M}{\beta_{i}\Phi(B)x_{t,i}}\right)\boldsymbol{e_{t,p}} \\
  \Psi_{3,t} &=& \left[\left(\Phi(B)y_{t}-\sum_{i=1}^{M}{\beta_{i}\Phi(B)x_{t,i}}\right)^2-\sigma^{2}\right].
\end{eqnarray*}

\noindent Substituting these values  of $\pi_{t}$ into equation (\ref{17}) we   find that

\begin{eqnarray}\label{23}
    l(\boldsymbol{\lambda, \beta, \phi},\sigma^{2})&=&\sum_{t=p+1}^{N}{\log \pi_{t}} \nonumber\\
    &=&-\sum_{t=p+1}^{N}\log\left((N-p)+\boldsymbol\lambda_{1}^{T}\boldsymbol\Psi_{1,t}+\boldsymbol\lambda_{2}^{T}\boldsymbol\Psi_{2,t}+\lambda_{3}\Psi_{3,t}\right)
\end{eqnarray}

\vspace{.1in}

\noindent Since $0<\pi_{t}<1$,  the EL method maximizes  this function over the  set $(N-p)+\boldsymbol\lambda_{1}^{T}\boldsymbol\Psi_{1,t}+\boldsymbol\lambda_{2}^{T}\boldsymbol\Psi_{2,t}+\lambda_{3}\Psi_{3,t}>1$.   The  rest  of  this  optimization  problem will  be  carried  on  as  follows.

\vspace{.1in}

\noindent For  given $\boldsymbol{\beta}$,  $\boldsymbol{\phi}$  and $\sigma^{2}$  minimize the  function  $l(\boldsymbol{\lambda, \beta, \phi},\sigma^{2})$ given  in  equation (\ref{23}) with  respect to  the Lagrange multipliers  $\boldsymbol{\lambda}=(\boldsymbol{\lambda}_1,\boldsymbol{\lambda}_2,\lambda_3)$.  That  is, for  given $\boldsymbol{\beta}$,  $\boldsymbol{\phi}$  and $\sigma^{2}$ solve  the  following minimization  problem  to get the  values  of  Lagrange  multipliers

\begin{eqnarray*}
  \boldsymbol{\lambda}(\boldsymbol{\beta},\boldsymbol{\phi},\sigma^{2}) = argmin_{\boldsymbol{\lambda}}l(\boldsymbol{\lambda, \beta, \phi},\sigma^{2}).
\end{eqnarray*}

\vspace{.1in}

\noindent Since the   solution of  this  minimization problem cannot be obtained explicitly   numerical  methods  should  be  used to  get     solutions.  Substituting  this  solution  in    $l(\boldsymbol{\lambda, \beta, \phi},\sigma^{2})$  yields the   function  $l(\boldsymbol{\lambda}(\boldsymbol{\beta},\boldsymbol{\phi},\sigma^{2}),\boldsymbol{ \beta, \phi},\sigma^{2})$ that is only depend on   the  model  parameters $\boldsymbol{\beta}$,  $\boldsymbol{\phi}$  and $\sigma^{2}$.  This  function   can  be  regarded   as a profile conditional  empirical log-likelihood function.  The  CEL  estimators  $\hat{\boldsymbol{\beta}}$,  $\hat{\boldsymbol{\phi}}$  and $\hat{\sigma}^{2}$    will  be   obtained  by  maximizing $l(\boldsymbol{\lambda}(\boldsymbol{\beta},\boldsymbol{\phi},\sigma^{2}),\boldsymbol{ \beta, \phi},\sigma^{2})$   function  with  respect  to   the model  parameters $\boldsymbol{\beta}$,  $\boldsymbol{\phi}$  and $\sigma^{2}$.   That  is,  the  CEL  estimators  will  be   the  solutions  of  the  following  maximization problem
\begin{eqnarray*}
(\hat{\boldsymbol{\beta}},\hat{\boldsymbol{\phi}},\hat{\sigma}^{2})=argmax_{\boldsymbol{\beta},\boldsymbol{\phi},\sigma^{2}}l(\boldsymbol{\lambda}(\boldsymbol{\beta},\boldsymbol{\phi},\sigma^{2}),\boldsymbol{ \beta, \phi},\sigma^{2}).
\end{eqnarray*}

Since, there is no explicit solution of  this  maximization   problem  to  explicitly  obtain the  CEL  estimators $\hat{\boldsymbol{\beta}}$,  $\hat{\boldsymbol{\phi}}$  and $\hat{\sigma}^{2}$  numerical methods should be used to obtain CEL  estimators. In  this  paper, we use a Newton
type algorithm to carry  on  this optimization  problem.  The  steps  of our algorithm are as  follows.

\bigskip

\noindent \textbf{Step 0}. Set  initial values $\boldsymbol{\lambda }%
^{(0)}$, $\boldsymbol{\beta }^{(0)}$, $\boldsymbol{\phi }^{(0)}$ and $\sigma
^{2(0)}$. Fix  stopping rule $\epsilon $. The starting value of $\boldsymbol{%
\lambda }^{(0)}$ can be set to be the zero vector, but setting it $-n$ for
each lagrange multiplier yields  faster convergence.

\bigskip

\noindent \textbf{Step 1}. The function $l(\boldsymbol{\lambda ,\beta }^{(0)}%
\boldsymbol{,\phi }^{(0)},\sigma ^{2(0)})$ given in  equation (\ref{23}) is minimized with
respect to $\boldsymbol{\lambda }$.  This will be  done by  iterating

\begin{eqnarray*}
\boldsymbol{\lambda _{j+1}}=\boldsymbol{\lambda }_{j}+\frac{1}{2}(l_{%
\boldsymbol{\lambda }}^{\prime \prime -1}l_{\boldsymbol{\lambda }}^{\prime })
\end{eqnarray*}

\noindent until convergence  is satisfied. Here, $l_{\boldsymbol{\lambda }}^{\prime }$ is the first order derivative
and $l_{\boldsymbol{\lambda }}^{\prime \prime }$ is the second order
derivative of $l(\boldsymbol{\lambda ,\beta }^{(0)}\boldsymbol{,\phi }%
^{(0)},\sigma ^{2(0)})$ with respect to $\boldsymbol{\lambda }$. By doing so
we calculate $\boldsymbol{\lambda }^{(m)} $  for $m=1,2,3,...$ value for the lagrange multipliers computed at $m-th$ step.

\bigskip

\noindent \textbf{Step 2}. After finding  $\boldsymbol{\lambda }^{(m)}
$ at Step 1, the  function $l(\boldsymbol{\lambda }^{(m)}\boldsymbol{,\beta ,\phi ,}\sigma^2
)$ is maximized with respect to $\boldsymbol{\phi} $, $\boldsymbol{\beta} $ and $\sigma^2 $ using
following updating equations.

\begin{eqnarray*}
\boldsymbol{\phi }_{j+1}=\boldsymbol{\phi }_{j}-\frac{1}{2}(l_{\phi
}^{\prime \prime -1}l_{\phi }^{\prime })
\end{eqnarray*}

\begin{eqnarray*}
\boldsymbol{\beta }_{j+1}=\boldsymbol{\beta }_{j}-\frac{1}{2}(l_{\beta
}^{\prime \prime -1}l_{\beta }^{\prime }).
\end{eqnarray*}

\begin{eqnarray*}
\sigma^{2}_{j+1}=\sigma^{2}_{j}-\frac{1}{2}(l_{\sigma^2 }^{\prime \prime -1}l_{\sigma^2
}^{\prime }).
\end{eqnarray*}

\noindent Here $l_{\beta }^{\prime }$, $l_{\phi }^{\prime }$ and $l_{\sigma^2}^{\prime }$  are the
first order derivatives of $l(\boldsymbol{\lambda }^{(m)},\boldsymbol{\beta
,\phi ,}\sigma^2 )$ with respect to $\boldsymbol{\beta }$, $\boldsymbol{\phi }$
and $\sigma^2 $ while $l_{\beta }^{\prime \prime }$, $l_{\phi }^{\prime
\prime }$ and $l_{\sigma^2 }^{\prime \prime}$ denote  the second order derivatives. At this step we
obtain $\boldsymbol{\beta }^{(m)}$, $\boldsymbol{\phi }^{(m)}$ and $\sigma
^{2(m)}$,  for $m=1,2,3,...$

\bigskip

\noindent \textbf{Step 3}. Together, steps 1 and 2 accomplish one iteration. Therefore, after steps 1 and 2 we obtain the values $\boldsymbol{\lambda }^{(m)}$, $\boldsymbol{\beta }^{(m)}$, $\boldsymbol{\phi }^{(m)}$ and $\sigma
^{2(m)}$ computed at  $m-th$ iteration. These are current estimates for the parameters and are used as initial values to obtain the estimates in $(m+1)-th$ iteration. Therefore, repeat steps 1 and 2 until


\begin{eqnarray*}
\left\Vert\begin{array}{c}
            \boldsymbol{\beta }^{(m+1)}-\boldsymbol{\beta }^{(m)}\\
            \boldsymbol{\phi }^{(m+1)}-\boldsymbol{\phi }^{(m)} \\
            \sigma ^{2(m+1)}-\sigma ^{2(m)}
          \end{array}\right\Vert<\epsilon
\end{eqnarray*}
is satisfied.

\bigskip

\noindent The performance of the proposed algorithm evaluated by the help of
simulation study and a numerical example given in the next section.

\section{Simulation Study}
\label{Sec3}

In this section, we give a small simulation study and a numerical example to illustrate the performance of the empirical likelihood estimators for the autoregressive error term regression model over the estimators obtained from the normally distributed error terms. We use R version 3.4.0 (2017-04-21) \cite{R2017} to carry  on  our    simulation study and  numerical example.

\subsection{Simulation Design}
\label{SubSec3.1}

We consider the second order (AR(2)) autoregressive error term model with M=3  regression parameters. The sample sizes are taken as $n=25,50$ and $100$.  For each case the explanatory variables  $\boldsymbol{x_t}$ are generated from standard normal distribution $(\boldsymbol{x_t}\sim N(0,1))$. The values of the parameters are taken as $\boldsymbol{\beta}=(\beta_1,\beta_2,\beta_3 )'  =(1, 3, 5)'$ ,$\sigma^2 = 1$ and $\boldsymbol{\phi}=(\phi_1,\phi_2 )'=(0.8,-0.2)' $. Note that the  values of $\boldsymbol{\phi}$ are taken as above to guarantee the stationarity assumption  of the error terms. We  consider  three   different  distributions for  the error term $a_t$:    $N(0,1)$, $0.9N(0,1)+0.1N(30,1)$  and $0.9N(0,1)+0.1N(0,10)$.   Note  that last  two    distributions  are used to generate outliers in y-direction.   After setting    the  values of  the  model   parameters $\boldsymbol{\beta}$,  $\boldsymbol{\phi}$  and $\sigma^{2}$  and deciding the distribution of  the error term, the values of the response variable are generated using $\phi(B) y_t =\sum_{i=1}^{M} \beta_i \phi(B) x_{t,i}+a_t$.\\

We  compare the  results of  the  CEL  estimators  with  the  results  of  the  CML  estimators. Mean squared error (MSE) and bias values are calculated as performance measures to compare the estimators.   These values are calculated  using the following equations for 1000 replications:

\bigskip

\noindent $MSE(\hat{\boldsymbol{\beta}})=(\frac{1}{1000} )\sum_{i=1}^{1000}(\hat{\boldsymbol{\beta}}_i-\boldsymbol{\beta})^2$, $bias(\hat{\boldsymbol{\beta}})=\overline{\boldsymbol{\beta}}-\boldsymbol{\beta}$,\\

\noindent$MSE(\hat{\boldsymbol{\phi}})=(\frac{1}{1000}) \sum_{i=1}^{1000}(\hat{\boldsymbol{\phi}_i}-\boldsymbol{\phi})^2$,  $bias(\hat{\boldsymbol{\phi}})=\overline{\boldsymbol{\phi}}-\boldsymbol{\phi}$,\\

\noindent $MSE(\hat{\sigma^2})=(\frac{1}{1000}) \sum_{i=1}^{1000}(\hat{\sigma}^2_i-\sigma^2)^2$, $bias(\hat{\sigma^2})=\overline{\sigma}^2-\sigma^2$\\

\noindent where  $\overline{\boldsymbol{\beta}}=(\frac{1}{1000} )\sum_{i=1}^{1000}\hat{\boldsymbol{\beta}}_i$  , $\overline{\boldsymbol{\phi}}=(\frac{1}{1000}) \sum_{i=1}^{1000}\hat{\boldsymbol{\phi}}_i$,and ${\overline{\sigma}^2}=(\frac{1}{1000}) \sum_{i=1}^{1000}\hat{\sigma}^2_i$.

\subsection{Simulation Results}
\label{SubSec3.2}

In Table \ref{T1}, mean estimates,  MSE and bias values  of  the CEL and  CML estimators  computed  over  1000 replications    for  the  normally   distributed  error  case  (for  the  case  without outliers).  We  observe  from  these  results   that  the  both   estimators  have  similar  behavior  for the  large  sample  sizes  when   outliers are  not  the  case.  On the  other  hand,  for smaller  sample  sizes   the  CEL  estimators have better performance  than  the  CML  estimators  in terms  of  the  MSE values.  Thus,  for  small  sample  sizes the  estimator  based  on EL  method  may be  a  good  alternative  to  the  CML   method.

The  simulation  results   for  the  outlier cases are   reported   in  Table \ref{T2} and \ref{T3}.   From these tables
we observe that when outliers are introduced in the data the CML estimator  is noticeably influenced by the outliers with higher  MSE values. On the  other hand, the estimators based on empirical likelihood still retain their good performance in terms of  having  smaller  MSE and bias values. To sum up, the performance of the estimators obtained from the empirical likelihood is superior to the performance of the estimators obtained from CML method  when some outliers are present in the data  and/or   the  sample  size  is  smaller.

\subsection{Numerical Example}
\label{SubSec3.3}
Montgomery, Peck, and Vining \cite{Montgomery2012} give soft drink data example relating annual regional advertising expenses to annual regional concentrate sales for a soft drink company for 20 years. After calculating LS residuals for the linear regression model, the assumption of uncorrelated errors can be tested by using the Durbin--Watson statistic, which is a test statistics detecting the positive autocorrelation. The critical values of the Durbin--Watson statistic are dL = 1.20 and dU = 1.41 for one explanatory variable and 20 observations. The calculated value of the Durbin--Watson statistic d = 1.08 and this value is less than the critical value dL = 1.20. Therefore it can be said that the errors in the regression model are positively autocorrelated.

\begin{figure}
\centering
\subfloat[without outlier]{%
\resizebox*{7cm}{!}{\includegraphics{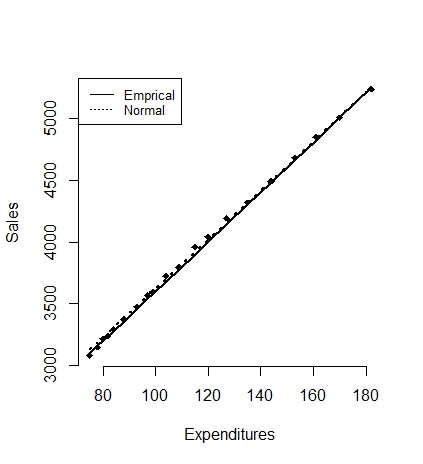}}}\hspace{5pt}
\subfloat[with one outlier]{%
\resizebox*{7cm}{!}{\includegraphics{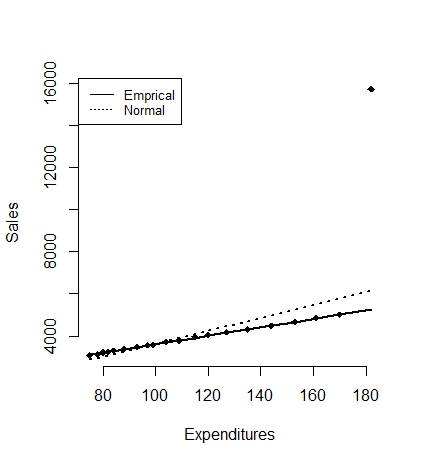}}}
\caption{Scatter plot of soft drink data and, CEL and CML fits with outlier} \label{Fig1}
\end{figure}

We  consider a linear regression  model  with  AR(p) error  terms  and   find   the  LS  estimators for  the  model parameters. Then, we use the LS estimators   as the initial values to  run  the  algorithm to  compute CEL and CML  estimators.  The estimates of  the  parameters  are given in Table \ref{T4}. We  observe  from the results given in  Table \ref{T4} that  the CEL and the CML methods give   similar estimates for the regression and the autoregressive parameters.  But,  the CEL estimate of error  variance  is smaller than the  estimate obtained from the  CML method. Figure \ref{Fig1}(a) shows the scatter plot  of the data set with
the fitted regression  lines  obtained  from  CEL  and CML  methods.  We observe
that the fitted lines  are  coinciding.  To  further  explore  the behavior of  the  estimators   against  the outliers we create one outlier  by  replacing  the last  observation  with  an  outlier. For  the new  data set (the  same  data set  with  one  outlier)   we again   find the CEL  and  the CML  estimates  for  all  the  parameters. We also provide these  estimates in Table \ref{T4}.  Figure \ref{Fig1}(b) depicts the scatter plot of the data and the fitted lines obtained from the CEL  and the CML  estimates. We observe  that, unlike  the  fitted  line  obtained  from  the CEL  method, the fitted line  obtained  from the  CML method  is badly influenced by the outlier.

\subsection{Real Data Example}
\label{SubSec3.4}
\begin{figure}
\centering
\subfloat[without outlier]{%
\resizebox*{7cm}{!}{\includegraphics{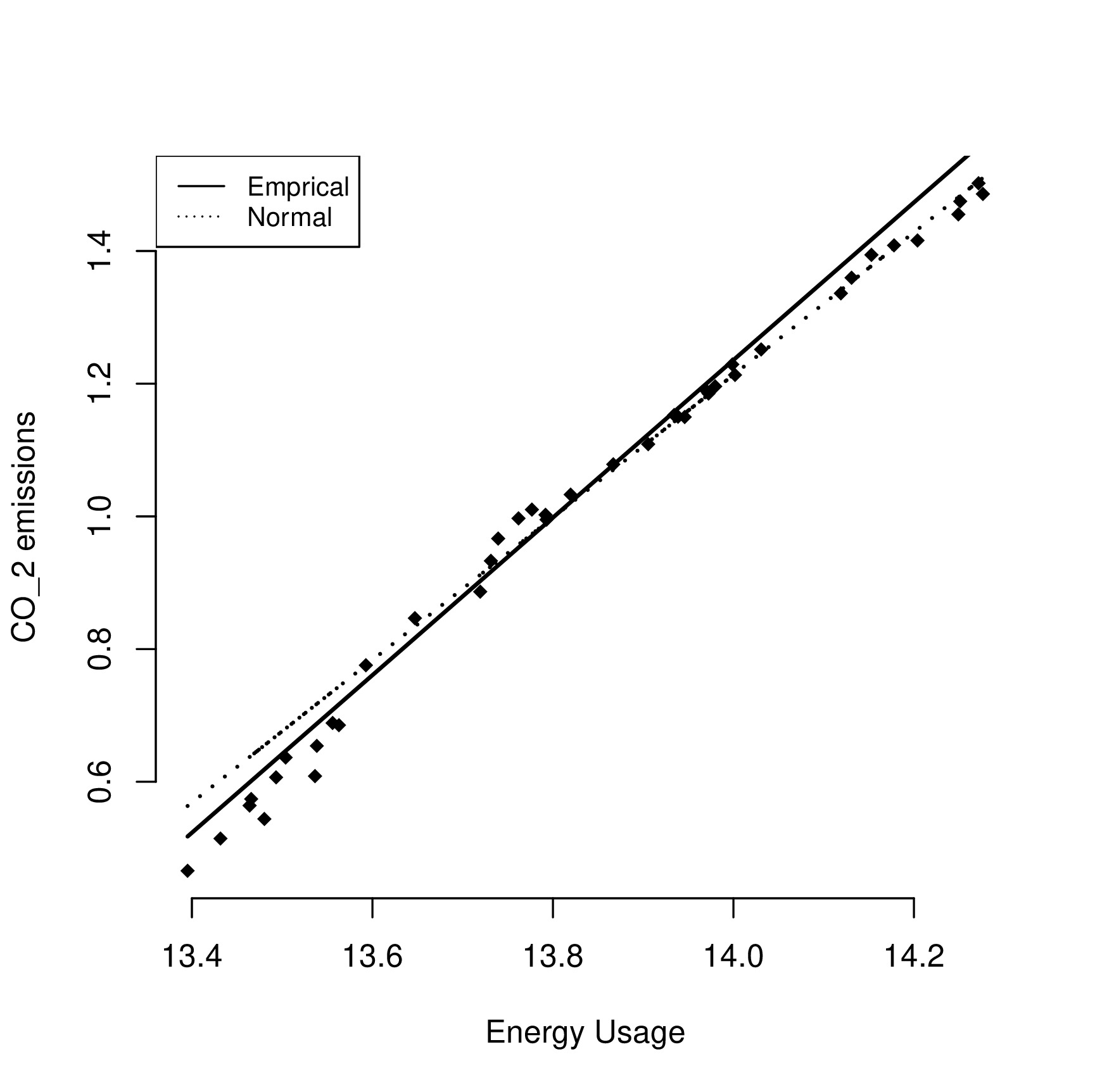}}}\hspace{5pt}
\subfloat[with one outlier]{%
\resizebox*{7cm}{!}{\includegraphics{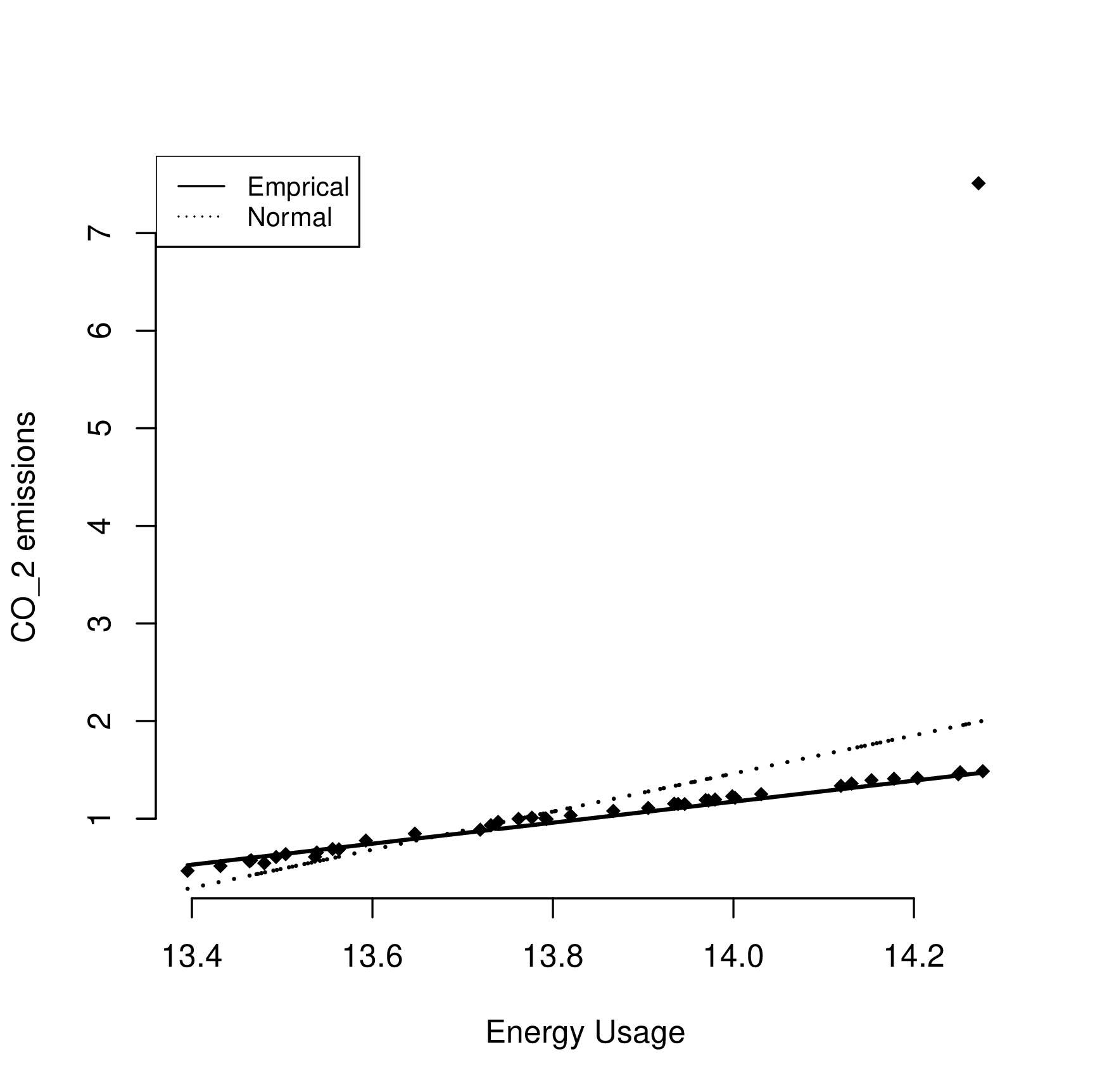}}}
\caption{Scatter plot of $CO_2$ emission- Energy usage data,and CEL and CML fits} \label{Fig2}
\end{figure}

\begin{figure}
  \includegraphics[width=1\linewidth]{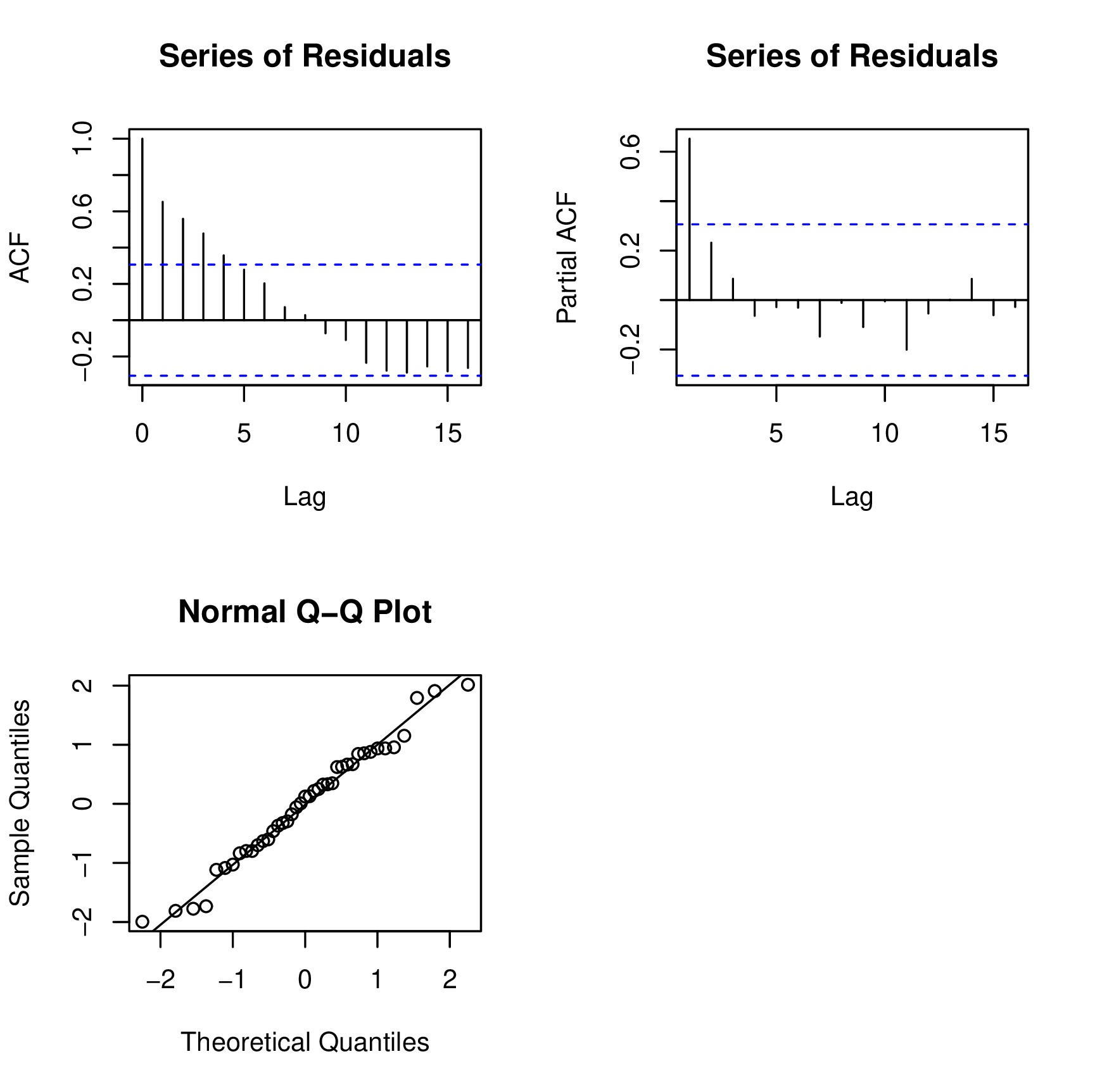}\\
  \caption{ACF, PACF and Q-Q of plots of residuals $CO_2$ emission- Energy usage data}
  \label{Fig3}
\end{figure}

We use a real data set consisting of $CO_2$ emission (metric kgs per capita) and energy usage (kg of oil equivalent per capita) of Turkey between the years 1974-2014. The data set can be obtained from ``https://data.worldbank. org". The scatter plot in Figure \ref{Fig2}(a) show that a linear relationship between logarithms of  $CO_2$ emission and energy usage. If the error terms are calculated using the LS estimates (-15.06489, 1.16258), it can be seen that the data has  AR(1) structure with $\hat\phi=0.65$ autocorrelation coefficient, since the Durbin-Watson test statistic and its p-value are 0.662571 and 1.235e-07,respectively. The ACF (autocorrelation function) and PACF (Partial ACF) plots given in Figure \ref{Fig3} also support the same results. Further, Q-Q plot of residuals in Figure \ref{Fig3} yield that the assumption of normally distributed error terms is achieved.

For this data set,we use the CML and CEL estimates to estimate the parameters of regression model and autocorrelation structure and also error variance.  Note that the LS estimates of original data are used as initial values to compute the CML and the CEL estimates given in Table \ref{T5}. The fitted regression lines along with the scatter plot of original data are also given in Figure \ref{Fig2}(a),and we can see that both methods give similar results.

To evaluate the performance of CEL method when there are outliers in the data or departure from normality, we  create an artificial outlier  multiplying last observation by five in the data, and use the both estimation methods to estimate the model parameters. The estimates are provided in Table \ref{T5} and the fitted lines obtained from both methods can be seen in Figure \ref{Fig2}(b). We can easily see from Figure \ref{Fig2}(b), the CEL method have better fit than the CML method to the data with outlier. Although real autocorrelation is positive, the CML estimate of autocorrelation is calculated as negative because of the outlier effect. In this case the LS estimates of regression parameters are (-27.4952, 2.0717) and estimated autocorrelation is -0.0107.

\section{Conclusion}
\label{Sec4}

In  literature, the  CML  or the LS  methods  are  often  adapted  to  estimate the  parameters of an  autoregressive  error  terms  regression  model. The  CML  method are  applied  to  carry on  the  estimation  under  some  distributional  assumption  on the  error   therm.  However,  for  some  data sets  it  may  not   be plausible   to  make  some  distributional  assumption  on  the  error  term  due  to  the  lack  of  information  about  data sets.   For those  data sets,  some  alternative  distributional  assumption  free methods  should  be  preferred to   continue regression   analysis. One  of   those  distributional  free   estimation  methods  is  the EL  method,  which  can also  be  used for small  sample  sizes.  In  this  paper,  we  have used  the EL  estimation  method  to  estimate   the  parameters  of an autoregressive  error  terms  regression  model.  We   have  defined  a  CEL function and constructed the   necessary  constraints using  probability  weights and the  normal  equations borrowed  from the  classical LS estimation  method to  conduct   parameter  estimation.  To  evaluate and  compare  the  performance  of  the proposed   method  with the CML  method  we  have  provided  a   simulation  study   and an example. We  have   compared  the results using the MSE and  the bias. We  have  designed two different  simulation  scenarios,  data  with  outliers  and  data  without  outliers.    The results of simulation study, numerical and real data examples have demonstrated that the CEL method can perform better than  the CML   method when   there   are  outliers in  the  data set, which symbolizes the deviation from normality assumption.  On  the  other  hand,  they  have similar  performance if  there   are  no  outliers in the  data sets.\\

\newpage

\begin{table}[h]
\tbl{Estimates, MSE and Bias values of the estimates from $a_t \sim N(0,1)$ for different sample sizes without outliers. True values are $(\beta_1,\beta_2,\beta_3 )'  =(1, 3, 5)'$,  $(\phi_1,\phi_2 )'=(0.8,-0.2)$ and $\sigma^2=1$}%
{\begin{tabular}{llrrrrrr}\toprule
                  && \multicolumn{2}{c}{n=25} & \multicolumn{2}{c}{n=50}& \multicolumn{2}{c}{n=100}  \\ \cmidrule{3-8}
                  &&Normal&Empirical&Normal&Empirical& Normal&Empirical   \\ \midrule
\rowcolor[HTML]{EFEFEF}
\multirow{3}{*}{\cellcolor[HTML]{EFEFEF}} &$\hat{\beta}_{1}$
       &0.98796 &1.00138 &0.99845 &1.00178 &0.99932 &1.00068\\
\rowcolor[HTML]{EFEFEF}
\cellcolor[HTML]{EFEFEF}{$\beta_{1}$}
&MSE   &0.03276 &0.0058 &0.01445 &0.00228 &0.00584 &0.00074\\
\rowcolor[HTML]{EFEFEF}
\cellcolor[HTML]{EFEFEF}
&BIAS  &0.14255 &0.05276 &0.09522 &0.0319 &0.0605 &0.01832\\
\multirow{3}{*}{$\beta_{2}$} &$\hat{\beta}_{2}$
        &3.00136 &3.00697 &2.99808 &3.00009 &2.9978  &2.99753\\
&MSE    &0.03247 &0.00613 &0.01346 &0.00225 &0.00664 &0.00071\\
&BIAS   &0.14406 &0.05512 &0.09181 &0.03214 &0.06437 &0.01797\\
\rowcolor[HTML]{EFEFEF}
\multirow{3}{*}{\cellcolor[HTML]{EFEFEF}} &$\hat{\beta}_{3}$
        &5.00803 &5.00139 &4.99681 &4.99966 &4.99913 &5.00266\\
\rowcolor[HTML]{EFEFEF}
\cellcolor[HTML]{EFEFEF}{$\beta_{3}$}
&MSE    &0.03481 &0.00657 &0.01421 &0.00202 &0.00583 &0.00069\\
\rowcolor[HTML]{EFEFEF}
\cellcolor[HTML]{EFEFEF}
&BIAS   &0.14491 &0.05574 &0.09389 &0.03042 &0.06045 &0.01789\\
\multirow{3}{*}{$\phi_{1}$} &$\hat{\phi}_{1}$
       &0.76906 &0.80372 &0.77997 &0.8013  &0.79587	&0.80081\\
&MSE   &0.0457  &0.00028 &0.02118 &0.0001  &0.01042	&0.00004\\
&BIAS  &0.16869	&0.01328 &0.11537 &0.00784 &0.08196	&0.00465\\
\rowcolor[HTML]{EFEFEF}
\multirow{3}{*} {\cellcolor[HTML]{EFEFEF}} &$\hat{\phi}_{2}$
       &-0.2095 &-0.19714 &-0.20945 &-0.199 &-0.21079 &-0.20002\\
\rowcolor[HTML]{EFEFEF}
\cellcolor[HTML]{EFEFEF}{$\phi_{2}$}
&MSE   &0.04266	&0.00029 &0.01949 &0.0001 &0.00951 &0.00004\\
\rowcolor[HTML]{EFEFEF}
\cellcolor[HTML]{EFEFEF}
&BIAS  &0.16559	&0.01394 &0.11067 &0.00824 &0.07779 &0.00467\\
\multirow{3}{*}{$\sigma^2$}    &$\hat{\sigma}^{2}$
      &0.88229 &1.15221 &0.94378 &1.1022 &0.96983 &1.07259\\
&MSE  &0.03511 &0.03062 &0.01351 &0.01454&0.00633 &0.00784\\
&BIAS &0.15368 &0.15458 &0.09334 &0.1022 &0.06461 &0.07326\\
                             \bottomrule
\end{tabular}}
\label{T1}
\end{table}

\begin{table}[h]
\tbl{Estimates, MSE and Bias values of the estimates for different sample sizes from $a_t\sim 0.9N(0,1)+0.1N(30,1)$. True values are $(\beta_1,\beta_2,\beta_3 )'  =(1, 3, 5)'$,  $(\phi_1,\phi_2 )'=(0.8,-0.2)$ and $\sigma^2=1$}%
{\begin{tabular}{llrrrrrr}\toprule
                  && \multicolumn{2}{c}{n=25} & \multicolumn{2}{c}{n=50}& \multicolumn{2}{c}{n=100}  \\ \cmidrule{3-8}
                  &&Normal&Empirical&Normal&Empirical& Normal&Empirical   \\ \midrule
\rowcolor[HTML]{EFEFEF}
\multirow{3}{*}{\cellcolor[HTML]{EFEFEF}} &$\hat{\beta}_{1}$
        &0.99758 &1.0001 &1.01599 &1.0008 &0.9947 &1.00067\\
\rowcolor[HTML]{EFEFEF}
\cellcolor[HTML]{EFEFEF}{$\beta_{1}$}
&MSE    &0.53329 &0.00588 &0.11102&0.00189&0.02944 &0.00071\\
\rowcolor[HTML]{EFEFEF}
\cellcolor[HTML]{EFEFEF}
&BIAS   &0.56527 &0.05219 &0.26139&0.02957&0.13598 &0.01776\\
\multirow{3}{*}{$\beta_{2}$} &$\hat{\beta}_{2}$
        &2.97227 &3.00938 &2.98684 &2.9988  &2.99961 &3.00186\\
&MSE    &0.51638 &0.00659 &0.10983 &0.00222 &0.02613 &0.00068\\
&BIAS   &0.5623  &0.05636 &0.26129 &0.03185 &0.12822 &0.01766\\
\rowcolor[HTML]{EFEFEF}
\multirow{3}{*}{\cellcolor[HTML]{EFEFEF}} &$\hat{\beta}_{3}$
        &4.97749 &5.00772 &5.00388 &5.00511 &4.99569 &5.00152\\
\rowcolor[HTML]{EFEFEF}
\cellcolor[HTML]{EFEFEF}{$\beta_{3}$}
&MSE    &0.50668 &0.00559 &0.10949 &0.00218 &0.0282 &0.00072\\
\rowcolor[HTML]{EFEFEF}
\cellcolor[HTML]{EFEFEF}
&BIAS   &0.55812 &0.05204 &0.2574 &0.03174 &0.13337 &0.01798\\
\multirow{3}{*}{$\phi_{1}$}  &$\hat{\phi}_{1}$
        &1.77218 &0.80245 &1.7009  &0.80143 &1.63445 &0.80104\\
&MSE    &0.9494  &0.00029 &0.81299 &0.00009 &0.69691 &0.00003\\
&BIAS   &0.97218 &0.0136  &0.9009  &0.00746 &0.83445 &0.00446\\
\rowcolor[HTML]{EFEFEF}
\multirow{3}{*}{\cellcolor[HTML]{EFEFEF}} &$\hat{\phi}_{2}$
        &-0.94992 &-0.19518 &-0.75483 &-0.19879 &-0.64632 &-0.20002\\
\rowcolor[HTML]{EFEFEF}
\cellcolor[HTML]{EFEFEF}{$\phi_{2}$}
&MSE    &0.58956 &0.00033 &0.31046 &0.00011 &0.19993 &0.00004\\
\rowcolor[HTML]{EFEFEF}
\cellcolor[HTML]{EFEFEF}
&BIAS   &0.74992 &0.01471 &0.55483 &0.0083  &0.44632 &0.00462\\
\multirow{3}{*}{$\sigma^2$} &$\hat{\sigma}^{2}$
        &5.95214 &1.13923 &4.42534 &1.10108 &3.30909 &1.06915\\
&MSE    &24.68756&0.02581 &11.7763 &0.0142 	&5.34667 &0.00688\\
&BIAS   &4.95214 &0.1419  &3.42534 &0.10183 &2.30909 &0.06921\\
                             \bottomrule
\end{tabular}}
\label{T2}
\end{table}

\begin{table}[h]
\tbl{Estimates, MSE and Bias values of the estimates for different sample sizes from $a_t \sim 0.9N(0,1)+0.1N(0,10)$. True values are $(\beta_1,\beta_2,\beta_3 )'  =(1, 3, 5)'$,  $(\phi_1,\phi_2 )'=(0.8,-0.2)$ and $\sigma^2=1$}%
{\begin{tabular}{llrrrrrr}\midrule
                  && \multicolumn{2}{c}{n=25} & \multicolumn{2}{c}{n=50}& \multicolumn{2}{c}{n=100}  \\ \cmidrule{3-8}
                  &&Normal&Empirical&Normal&Empirical& Normal&Empirical   \\ \midrule
\rowcolor[HTML]{EFEFEF}
\multirow{3}{*}{\cellcolor[HTML]{EFEFEF}} &$\hat{\beta}_{1}$
         &1.00012 &1.00766 &0.9878 &0.99887 &0.99299 &1.00153 \\
\rowcolor[HTML]{EFEFEF}
\cellcolor[HTML]{EFEFEF}{$\beta_{1}$}
&MSE    &0.21859 &0.00777 &0.11943 &0.00276 &0.06829 &0.00076 \\
\rowcolor[HTML]{EFEFEF}
\cellcolor[HTML]{EFEFEF}
&BIAS   &0.33632 &0.05886 &0.26183 &0.034   &0.20502 &0.01859 \\
\multirow{3}{*}{$\beta_{2}$} &$\hat{\beta}_{2}$
        &2.99806 &2.9978  &2.98896 &3.00449 &2.98302 &3.00347 \\
&MSE    &0.23989 &0.0086  &0.13169 &0.00265 &0.06363 &0.00083 \\
&BIAS   &0.34908 &0.06087 &0.2692  &0.03391 &0.19512 &0.01922 \\
\rowcolor[HTML]{EFEFEF}
\multirow{3}{*}{\cellcolor[HTML]{EFEFEF}} &$\hat{\beta}_{3}$
        &5.01594 &5.00257 &4.99975 &5.00495 &4.99407 &4.99616 \\
\rowcolor[HTML]{EFEFEF}
\cellcolor[HTML]{EFEFEF}{$\beta_{3}$}
&MSE    &0.22748 &0.0079  &0.12487 &0.00273 &0.06541 &0.00083 \\
\rowcolor[HTML]{EFEFEF}
\cellcolor[HTML]{EFEFEF}
&BIAS   &0.3443 &0.05964  &0.26638 &0.03409 &0.19757 &0.01894 \\
\multirow{3}{*}{$\phi_{1}$}  &$\hat{\phi}_{1}$
        &0.79014 &0.80161 &0.73106 &0.80076 &0.7298  &0.80066\\
&MSE    &0.43476 &0.00022 &0.18784 &0.00009 &0.08888 &0.00004 \\
&BIAS   &0.51506 &0.01213 &0.34467 &0.00757 &0.23919 &0.00471 \\
\rowcolor[HTML]{EFEFEF}
\multirow{3}{*}{\cellcolor[HTML]{EFEFEF}} &$\hat{\phi}_{2}$
        &-0.46927 &-0.19645 &-0.26362 &-0.19819 &-0.21955 &-0.19966 \\
\rowcolor[HTML]{EFEFEF}
\cellcolor[HTML]{EFEFEF}{$\phi_{2}$}
&MSE    &0.7917 &0.00031 &0.23481 &0.0001 &0.08613 &0.00004 \\
\rowcolor[HTML]{EFEFEF}
\cellcolor[HTML]{EFEFEF}
&BIAS   &0.70074 &0.01421 &0.37436 &0.00786 &0.23868 &0.00473 \\
\multirow{3}{*}{$\sigma^2$}    &$\hat{\sigma}^{2}$
        &2.65719 &1.14735 &2.70776 &1.09725 &2.94881 &1.06939 \\
&MSE    &3.79346 &0.02846 &3.59029 &0.01368 &4.16932 &0.00702 \\
&BIAS   &1.65951 &0.14943 &1.70791 &0.09889 &1.94881 &0.0697 \\
                             \bottomrule
\end{tabular}}
\label{T3}
\end{table}

\begin{table}[h]
\tbl{The parameter estimates for soft drink data}%
{\begin{tabular}{lrrrr}\toprule
                  & \multicolumn{2}{c}{without outlier} & \multicolumn{2}{c}{with one outlier}  \\ \cmidrule{2-5}
                  &Empirical&Normal&Empirical&Normal   \\ \midrule
\rowcolor[HTML]{EFEFEF}{$\hat{\beta}_{0}$} & 1593.95911&1645.37912&1617.57068&560.9601 \\
{$\hat{\beta}_{1}$}                        &   20.07335&  19.80988&  19.93477& 30.7569 \\
\rowcolor[HTML]{EFEFEF}{$\hat{\phi}_{1}$}  &    0.89579&   0.56856&   0.89814& -2.5689\\
{$\hat{\sigma}^{2}$}                       &    6.43127&  16.96544&   2.44785& 2025.174\\
 \bottomrule
\end{tabular}}
\label{T4}
\end{table}

\begin{table}[h]
\tbl{The parameter estimates for $CO_2$ emission-Energy usage data}%
{\begin{tabular}{lrrrr}\toprule 
                  & \multicolumn{2}{c}{without outlier} & \multicolumn{2}{c}{with one outlier}  \\ \cmidrule{2-5}
                  &Empirical&Normal&Empirical&Normal   \\ \midrule 
\rowcolor[HTML]{EFEFEF}{$\hat{\beta}_{0}$} &-15.389316&-13.827491&-13.891223&-25.869536\\
{$\hat{\beta}_{1}$}                        &  1.187502&  1.074341&  1.076084&  1.952316\\
\rowcolor[HTML]{EFEFEF}{$\hat{\phi}_{1}$}  &  0.655439&  0.861551&  0.511249& -0.201119\\
{$\hat{\sigma}^{2}$}                       &  0.824727&  0.021447&  1.148087&  0.900177\\
 \bottomrule
\end{tabular}}
\label{T5}
\end{table}

\newpage

\end{document}